\documentclass[conference]{IEEEtran}
\IEEEoverridecommandlockouts
\UseRawInputEncoding
\usepackage{amsmath,amssymb,amsfonts}
\usepackage{algorithmic}
\usepackage{graphicx}
\usepackage{textcomp}
\usepackage{xcolor}
\usepackage[utf8]{inputenc}
\usepackage[T1]{fontenc} 
\usepackage{tcolorbox}
\usepackage{dirtytalk}
\usepackage{multirow}
\usepackage{booktabs}
\usepackage{tablefootnote}
\usepackage[official]{eurosym}
\usepackage[
backend=bibtex, style=ieee]{biblatex}
\usepackage[colorlinks=true,linkcolor=black,anchorcolor=black,citecolor=black,filecolor=black,menucolor=black,runcolor=black,urlcolor=black]{hyperref}

\usepackage{makecell}

\newif\ifdraft
\drafttrue  %switch to <<draftfalse>> to remove all comments
\ifdraft

\newcommand{\alnote}[1]{ {\textcolor{magenta} {**Andre: #1}}}
\newcommand{\kaja}[1]{ {\textcolor{cyan} {**Kaja: #1}}}
\newcommand{\gm}[1]{ {\textcolor{blue} {**Gonzalo: #1}}}
\newcommand{\chr}[1]{ {\textcolor{green} {**Christopher: #1}}}
\else
\newcommand{\alnote}[1]{}
\newcommand{\kaja}[1]{}
\newcommand{\gm}[1]{}
\newcommand{\chr}[1]{}
\fi

\begin{document}

\title{Exploring privacy-enhancing technologies in the automotive value chain\\

\thanks{\textsuperscript{*}Corresponding author}
}

\makeatletter
\newcommand{\linebreakand}{%
  \end{@IEEEauthorhalign}
  \hfill\mbox{}\par
  \mbox{}\hfill\begin{@IEEEauthorhalign}
}
\makeatother

\author{\IEEEauthorblockN{Gonzalo Munilla Garrido\textsuperscript{*}}
\IEEEauthorblockA{ \textit{Department of Informatics, TUM} \\
\textit{Group IT, BMW Group}\\
Munich, Germany \\
gonzalo.munilla-garrido@tum.de}
\and
\IEEEauthorblockN{Kaja Schmidt}
\IEEEauthorblockA{\textit{Department of Business Informatics} \\
\textit{University of Potsdam}\\
Potsdam, Germany \\
kaja.schmidt@uni-potsdam.de}
\and

\IEEEauthorblockN{Christopher Harth-Kitzerow}
\IEEEauthorblockA{\textit{Department of Informatics, TUM} \\
\textit{Group IT, BMW Group}\\
Munich, Germany \\
christopher.harth-kitzerow@tum.de}

\linebreakand

\IEEEauthorblockN{Johannes Klepsch}
\IEEEauthorblockA{\textit{Group IT, BMW Group}\\
Munich, Germany \\
Johannes.Klepsch@bmw.de}
\and

\IEEEauthorblockN{Andre Luckow}
\IEEEauthorblockA{\textit{Group IT, BMW Group}\\
Munich, Germany \\
Andre.Luckow@bmwgroup.com}
\and

\IEEEauthorblockN{Florian Matthes}
\IEEEauthorblockA{\textit{Department of Informatics} \\
\textit{Technical University of Munich}\\
Munich, Germany  \\
Matthes@in.tum.de}

}

\IEEEoverridecommandlockouts
\IEEEpubid{\makebox[\columnwidth]{978-1-6654-3902-2/21/\$31.00~\copyright2021 IEEE \hfill} \hspace{\columnsep}\makebox[\columnwidth]{ }}

\maketitle

\IEEEpubidadjcol

% is attracting heightened attention due to its potential benefits in industry, like building brand differentiation, new business models\alnote{What new business models for automotive are you thinking of?}, or establishing a competitive advantage, as well as the increasing number of data breaches and recent privacy regulations.
%  considerations and still see privacy as a requirement rather than an opportunity to leverage big data in new ways because of the strong regulatory enforcement.
% enabled by privacy-enhancing products that can potentially bring new business models and improvements to products and services.
% The results indicate that many privacy use cases are not necessarily business-focused but essential for business activities. 
% Furthermore, the results emphasize the importance of organizations becoming agile and confident with privacy-enhancing data management internally before endeavoring into external use cases such as data markets.
% \alnote{Based on what evidence/data do we come to that conclusion?} We conclude that institutions that do not invest in privacy-enhancing products will lose their competitive advantage, even more so in the field of analytics.

\begin{abstract}
 Privacy-enhancing technologies (PETs) are becoming increasingly crucial for addressing customer needs, security, privacy (e.\,g., enhancing anonymity and confidentiality), and regulatory requirements. However, applying PETs in organizations requires a precise understanding of use cases, technologies, and limitations. This paper investigates several industrial use cases, their characteristics, and the potential applicability of PETs to these. We conduct expert interviews to identify and classify uses cases, a gray literature review of relevant open-source PET tools, and discuss how the use case characteristics can be addressed using PETs' capabilities. While we focus mainly on automotive use cases, the results also apply to other use case domains.
\end{abstract}


\maketitle

\section{Introduction}
\label{sec:intro}

% WHY important? %brogi_qos-aware_2017,
Data, analytics, and artificial intelligence (AI) are playing an increasingly important role across the automotive value chain~\cite{ramezani_industry40,7363874,8622357, 2021_a_Eirich_et_al}. 
The capabilities of AI are catalyzed by the growing deployment and use of Internet-of-Things devices~\cite{montazerolghaem_load-balanced_2020}. 
However, as the number of applications grows, the need to utilize advanced privacy-enhancing technologies (PETs) to improve data privacy, security, trust, and regulatory compliance (e.\,g., the European General Data Protection Regulation (GDPR~\cite{GDPR} and the Consumer Privacy Act), is increasing~\cite{zoll2021privacy}.  
Thus, PETs must and will become a foundational pillar of modern data platforms~\cite{Yuming2021}.

In addition to mitigating privacy, reputational and financial risks~\cite{jaatun2012, data_breach}, the usage of PETs has many benefits for institutions: a careful deployment of PETs may \emph{increase} not only trust but also data usage and collection as PETs help to overcome customer concerns~\cite{kaaniche2016abs}. 
By doing so, PETs can accelerate existing processes and enable new business models~\cite{privacy_economic_good, data_exchange_pros}. 
An example is the ability to support cross-organizational collaboration and data exchanges using PETs that provide the necessary trust and security for widespread adoption.

% The name PETs was first coined to encapsulate technologies that took a novel approach in privacy potection~\cite{hes_privacy_enhancing_1998}.
The term PETs, initially coined in 1995~\cite{hes_privacy_enhancing_1998}, encapsulates technologies designed to protect personal and sensitive data in-use by minimizing their exposure to potential malicious entities. PETs are complementary to established data security practices, e.\,g., in-transit and at-rest encryption. To reduce exposure, PETs rely on different mechanisms (e.\,g., cryptography) to conceal the information (confidentiality) or modify data to perturb the link with the data owner (anonymity).
Prominent PETs that enhance anonymity are differential privacy (DP)~\cite{DP_definition} and k-anonymity~\cite{samarati_protecting_nodate}, while secure multi-party computation (SMC)~\cite{SMC_basics} or homomorphic encryption (HE)~\cite{HomomorphicPaper} focus on confidentiality.
While each PET contributes uniquely to enhancing privacy, employing them in combination provides more holistic protection.

Many automotive use cases with complex requirements can benefit from numerous privacy-enhancing technologies~\cite{data_makets_properties_I,garrido_revealing_2021,trask_structured_transparency_nodate}. However, understanding use cases characteristics and requirements and the capabilities of PETs are often challenging~\cite{za_privacy_sensitive_2021}. While much research focuses on the capabilities of specific PETs and use cases~\cite{bittner2021private,choudhury2020differential,arfaoui2014}, there is a gap in surveying and mapping use cases to the PETs landscape.

% Related work reduces the overall complexity by focusing on implementing single PETs to a specific use case~\cite{bittner2021private}\cite{choudhury2020differential}\cite{arfaoui2014}, surveying how PETs address privacy requirements in general~\cite{heurix2015}, or proposing use cases without mapping them precisely to PETs~\cite{gonczol2020}\cite{theissler2021}.

% However, few publications focus on providing a systematic mapping between industrial use cases and PETs so that practitioners can distinctly recognize the unique contribution of each PET in a use case.
% To tackle this research gap, our work provides a set of reference industry use cases precisely mapped to PETs based on a set of suitable capabilities.
% In this manner, practitioners can identify the capabilities for their use cases based on our work and map them to the corresponding PETs.
% Overall, our goal is to guide practitioners across the bridge between use cases and PETs.

\textbf{Contributions}. 
We provide a comprehensive analysis of different application domains and use cases from the automotive value chain and discuss what characteristics and aspects of these use cases that can benefit from PETs. For this purpose, we investigate eight application domains, ranging from recommender systems, computer vision to data analytics. Based on a high-level overview, we provide an in-depth discussion of selected use cases, investigating the suitability of specific PETs. We identify important characteristics and patterns that allow practitioners to categorize new use cases and aid in identifying suitable PETs.

% can apply our research to other industries, we conduct the study from an automotive domain perspective. 
% Overall, we underline the variety of use cases an institution can pursue, the critical role PETs play across application domains to enable use cases, and the importance of adopting PETs to remain competitive.

% select suitable capabilities these use cases require in a production setting, identify the most relevant PETs and their open-source tooling to ensure these capabilities, and provide a concise mapping of each PET to a reference automotive use case.

% Specifically, we provide a clustering of privacy-related use cases in Table~\ref{tab:use_cases}, and describe the suitable capabilities of these use cases in section~\ref{subsec:capabilities}.
% The privacy-enhancing technologies that can meet these capabilities are listed in Table~\ref{tab:priv_tools} together with their most relevant open-source tools, and Table~\ref{tab:capabilities} maps them to the capabilities.
% Following the framework of Fig.~\ref{fig:framework}, we used the aforementioned capabilities to map eight reference use cases to their corresponding PETs in Table~\ref{tab:use_cases_pets}.

% \gm{I added and modified the previous 2 paragraphs as per your feedback.}
% \newline 

The remaining of the paper is structured as follows:
We introduce our methodology in section~\ref{sec:methodology}. 
We continue with an analysis of use cases and PETs in sections~\ref{sec:use_cases} and~\ref{sec:pets}. 
We discuss related work in section~\ref{sec:related_work}, and conclude in section~\ref{sec:conclusion}.

% \section{Terminology}
% \label{sec:terminology}

% \alnote{citations to definition, e.g. the use case one sounds very similar to the top result form goolg}

% In this section, we define three important concepts:
% A \emph{tool} is a reusable implementation of an algorithm that abstracts the deployment of a specific technology, i.e., the user does not need to have expertise in the underlying technology for its use.
% For example, an open-source library that abstracts the use of differential privacy to aggregate data in a privacy-enhancing manner.
% A \emph{use case} is a particular situation where a tool could enable or directly leverage a business opportunity or improve an existing process, product, or service.
% For example, the mood of a driver (situation) could be categorized on real time to accordingly adapt the environment within the vehicle for a better customer experience (business opportunity) with the use of privacy-enhancing machine learning (tool).
% A \emph{suitable capability} refers to a possible requirement a use case needs to fulfill in a productive setting, while a privacy-enhancing technology can provide such \emph{capability}.\gm{I refactored the last sentence}

\begin{table*}[t!]
  \centering
  \scriptsize
  \caption{Selected application domains and use cases.}
  \label{tab:use_cases}
  \begin{tabular}{|p{0.05cm}|p{1.5cm}|p{5cm}|p{9.5cm}|}
  \hline 
    \textbf{\#} &\textbf{Application \newline Domain}  & \textbf{Use Case} & \textbf{Description} \\
    \hline 
    1 & Recommender \newline systems 
    & Vehicle personalization, eco-friendly driving
    & Personalizing in-vehicle experiences and features based on data from in-vehicle sensors using analytics and machine learning, e.\,g., recommendations for music and locations, seat heating activation and  supporting gamification features (such as eco-friendly driving)~\cite{DBLP:conf/vtc/ChinTP20}.   \\
    \hline 
    2 & Geoservices 
    & Charging, traffic prediction, frequent routes, parking, charging, refuelling, points-of-interest
    & Geoservices enhance the travel experience based on highly-sensitive location data.
    \\
    \hline 
    3 & Computer \newline vision 
    & Attentiveness detection, visual quality inspection during manufacturing~\cite{8622357} 
    & Driver attention monitoring using camera-based systems and other sensor for improving safety. Data collected from cameras in-vehicle and in manufacturing plants is highly sensitive and may contain personal data, requiring PETs to ensure privacy.
    \\
    \hline 
    4 & Sensitive data \newline management 
    & Automation of anonymization pipelines, prolongation of data storage/access 
    & Creating, streamlining, or automating anonymization pipelines to implement regulatory complicance, increase data security and reduces human-error.
    \\
    \hline     
    5 & Data analytics 
    & Group statistics include business KPIs, sales statistics, demographics 
    & Analytics is essential to understand all aspects of the business, e.\,g., customer preferences, sales, and manufacturing performance~\cite{DBLP:conf/bigdataconf/JeereddyKDWV19}. 
    However, such statistics released publicly or confidentially for research or collaborative projects between institutions can lead an adversary to re-identify individuals~\cite{anonymizing_health_data}.
    \\
    \hline

    6 & Asset search 
    & Tracking components across value chains
    & Support tracking, search and reconciliation of assets across organizations, e.\,g., locating vehicle components in a supply chain~\cite{partchain}. To mitigate the risks of sharing data, data needs to be carefully curated and secured, preventing the sharing of sensitive information.
    \\
    \hline
    
    7 & 
    IoT 
    &  Connected vehicles %, connected IoT %, classified information processing
    & IoT deployments (vehicle, machines, etc.) produce vast amounts of data from on-board sensors and traffic infrastructure~\cite{7938385}. Data can be highly sensitive (e.\,g., behavioral data). PETs can reduce the need to centralize data in clouds.
    % Furthermore, companies with classified information without extensive local storage or computation resources might expose their secrets by offloading processing to third party cloud providers.
    \\
    \hline 
    
    8 & Cross-organizational data sharing 
    & Logistics \& supply chain data, data markets, KPI comparisons (industry benchmarks)
    & Sharing data across organizations to improve analytics and machine learning models (e.\,g., supply chain management and automated driving). PETs remove risk of sharing and the disclosure of sensitive and personal information to non-intended recipients.
    \\
    \hline 
    % 8 & Advertisement
    % & Billboard advertisement
    % & Advertising in a privacy-enhancing manner can avoid price discrimination or unwarranted tracking. Furthermore, automakers can leverage demographics and geo-positioning data to propose advertisers' billboard locations without disclosing personal information.
    % \\
    % \hline 
  \end{tabular}
  
\end{table*}
\section{Methodology}
\label{sec:methodology}

We investigate two research questions (RQs).
We interviewed several experts to identify and characterize use cases in the domain of privacy (RQ1). Further, we conducted a gray literature review to identify open-source tools that implement privacy-enhancing technologies (RQ2).  \newline 

\textit{\textbf{RQ1.}} \textit{What are the relevant use cases for PETs in the automotive industry?}
To answer this RQ, we provide use cases to motivate practitioners to enhance privacy in their institutions (see section~\ref{sec:use_cases}).

To plan and conduct the interviews to answer this RQ, we followed guidelines from P.~Runeson and M.~H\"ost~\cite{runeson2009guidelines}.
Specifically, throughout the end of 2020 and during the first half of 2021, we interviewed $17$ interested practitioners who worked directly or indirectly in the automotive industry; all the participants focused on data or privacy management.
Seven of the interviews were conducted verbally, while the remaining ten were through email correspondence.
The confidentiality of their identities and answers were communicated before initiating the interviews, as well as the goal of this study and how their answer will be used.
The interviews were semi-structured~\cite{runeson2009guidelines}, i.\,e., while we initiated the conversation with a set of preliminary questions about their background and followed up with RQ1 to collect a list of use cases, we promoted further exploration of their ideas revolving around their use cases list.
We countered potential bias by ensuring that the experts came from different organizational units and institutions and summarized the findings before the conclusion of the interview to get feedback and avoid misinterpretation~\cite{runeson2009guidelines}.

Afterward, we aggregated application domains and over $20$ use cases (see Table~\ref{tab:use_cases}). 
Based on the identified use cases, we identified characteristics that can be addressed by specific capabilities of available PETs to guide their implementation in a production setting: \textit{privacy}, \textit{function types}, \textit{data volume}, \textit{data authenticity}, \textit{query type}, and \textit{the number of interacting parties}.
Furthermore, we designed the framework of Fig.~\ref{fig:framework} to help us map in Table~\ref{tab:use_cases_pets} selected reference use cases to privacy-enhancing technologies.
\newline 

% Note that practitioners can also apply these use cases to other industries other than the automotive industry (e.g., geoservices based on points-of-interest or sensitive data management). \newline 

\textit{\textbf{RQ2.}} \textit{What are relevant privacy-enhancing tools available?}
During June and November 2021, we searched for tools practitioners can use to implement PETs in their use cases (see section~\ref{sec:pets}). We define a \emph{tool} as a reusable implementation of an algorithm that abstracts the deployment of a specific technology, i.e., the user does not need to have expertise in the underlying technology for its use.

We chose PETs included in seminal surveys or implementations in the domain of privacy~\cite{data_makets_properties_I,garrido_revealing_2021,trask_structured_transparency_nodate}.
We list the tools in Table~\ref{tab:priv_tools}.
Furthermore, each tool had to be open-source so that the scientific and engineering community could audit and freely access them.
However, systematically collecting peer-reviewed publications would not capture all the novel tools available~\cite{Hopewell2006}. 
Thus, for our purposes, S.~Hopewell and M. Clarke and S. Mallett~\cite{Hopewell2006}, and J.~Vom~Brocke et.~al~\cite{VomBrocke2009} indicated that a gray literature review would be a more optimal strategy.
Consequently, we included tools that appeared within the first $100$ Google search results for the search string ``\textit{PET name} AND \textit{open-source} AND \textit{tool} AND \textit{GitHub}''.
Two researchers searched independently (one identified $67$ tools while the other $63$), and merged the results into $76$ after removing duplicates ($52$). 
% Two researchers searched independently (one identified $48$ tools while the other $47$), and merged the results into $57$ after removing duplicates ($38$). 
% 48 + 19
% 47 + 16
% 57 + 19
% 38 + 14

% With RQ2, we investigate the evolving landscape of open-source tools that practitioners can leverage to deploy privacy-enhancing technologies (PETs) in their use cases 

\section{Applications in the Automotive Value Chain}
\label{sec:use_cases}

\begingroup
\setlength{\tabcolsep}{10pt} % Default value: 6pt
\renewcommand{\arraystretch}{1.2} % Default value: 1

\begin{table*}[t!]
  \centering
  \scriptsize
  \caption{Technologies and their most relevant open-source tools.}
  \label{tab:priv_tools}
  \begin{tabular}{|p{2.1cm}|p{6.8cm}|p{7.1cm}|}
  \hline 
    \textbf{Technology}  
    & \textbf{Description} 
    & \textbf{Tool} \\
    \hline 
    
    Differential \newline privacy (DP)  
    & Mathematically guarantees that the output of a dataset analysis is ``essentially'' identical, despite the presence or absence of an individual in the dataset~\cite{dwork_algorithmic_2013,DP_definition}. 
    & \href{https://www.github.com/google/differential-privacy}{Google-DP} (Python wrapper: \href{https://github.com/OpenMined/PyDP}{PyDP}), \href{https://www.github.com/opendp/smartnoise-core}{SmartNoise},
\href{https://github.com/IBM/differential-privacy-library}{diffprivlib}, \href{https://github.com/Quantalabs/DiffPriv}{DiffPriv}, 
\href{https://github.com/opendp}{OpenDP},
\href{https://github.com/dpcomp-org/dpcomp_core}{DPComp Core} and \href{https://github.com/uvm-plaid/chorus}{Chorus} (behind Uber's DP SQL). Focused on DP and deep learning: \href{https://github.com/tensorflow/privacy}{TensorFlow privacy} and \href{https://github.com/pytorch/opacus}{PyTorch Opacus}.
    \\
    \hline 
    
    K-anonymity
    & K-anonymity guarantees the indistinguishability of a record with k-1 number of others in a dataset~\cite{samarati_protecting_nodate}. K-anonymity is useful to anonymize datasets before usage.
    & \href{https://arx.deidentifier.org/}{ARX}, \href{https://amnesia.openaire.eu/}{Amnesia}, and \href{https://github.com/realrolfje/anonimatron}{Anonimatron}.
    \\
    \hline 
    
    Synthetic data 
    & Populate a synthetic dataset with the learned distribution of the real data by means of ML~\cite{dikici2020constrained,7004228}.
    & 
    \href{https://github.com/sdv-dev/SDV}{SDV},
    \href{https://paperswithcode.com/paper/zpy-open-source-synthetic-data-for-computer}{ZPY},
    \href{https://github.com/gretelai/gretel-synthetics}{Gretel},
    \href{https://github.com/getsynth/synth}{Synth},
    \href{https://github.com/ydataai/ydata-synthetic}{Ydata},
    \href{https://github.com/DataResponsibly/DataSynthesizer}{DataSynthesizer},
    \href{https://github.com/synthetichealth/synthea}{Synthea}, and
    \href{https://github.com/RealImpactAnalytics/trumania}{Trumania}.
    \\
    \hline 
    
    Zero-knowledge \newline proof (ZKP) 
    & Enables proof of authenticity of information without revealing or sharing the underlying data~\cite{first_introduced_ZKP,zkp_defs}. 
    & \href{https://github.com/xlab-si/emmy}{emmy},
\href{https://github.com/scipr-lab/dizk}{dizk},
\href{https://github.com/patractlabs/zkmega}{zkMega},
\href{https://github.com/scipr-lab/libsnark}{libsnark},
\href{https://github.com/scipr-lab/libiop}{libiop},
\href{https://github.com/PlasmNetwork/ZKRollups}{ZKRollups},
\href{https://github.com/ing-bank/zkrp}{ZKRP},
%\href{https://github.com/sec-bit/ckb-zkp}{ckb-zkp}, % Listed twice!
%\href{https://github.com/HorizenOfficial/ginger-lib}{ginger-lib}, % Listed twice!
%\href{https://github.com/0xProject/OpenZKP}{OpenZKP}, % Listed twice!
%\href{https://hackmd.io/@zkteam/gnark}{gnark}, % Listed twice!
\href{https://github.com/sec-bit/ckb-zkp}{ckb-zkp},
\href{https://github.com/HorizenOfficial/ginger-lib}{ginger-lib},
\href{https://github.com/0xProject/OpenZKP}{OpenZKP}, and
\href{https://hackmd.io/@zkteam/gnark}{gnark}.
    \\
    \hline     
    
    Secure multi-party computation (SMC)  
    & Parties can jointly compute a function without disclosing their inputs by employing secret sharing or garbled circuits~\cite{SMC_basics}.
    & \href{https://github.com/data61/MP-SPDZ}{Multi-Protocol SPDZ},
\href{https://github.com/cryptobiu/libscapi}{LIBSCAPI},
\href{https://github.com/lschoe/mpyc}{MPyC}, \href{https://github.com/facebookresearch/crypten}{CrypTen},
\href{https://github.com/emp-toolkit}{EMP-Toolkit},
\href{https://github.com/pillarjs/multiparty}{Multiparty}, 
\href{https://github.com/Zokrates/ZoKrates}{ZoKrates} and
\href{https://github.com/MPC-SoK/frameworks}{MPC-SoK}.
    \\
    \hline
   
    Homomorphic \newline  encryption (HE) 
    & Allows computing functions on ciphertext without prior decryption~\cite{HomomorphicPaper,Theory1}.
    & \href{https://github.com/tfhe/tfhe}{TFHE},
\href{https://github.com/IBM/fhe-toolkit-linux}{fhe-toolkit-linux},
\href{https://github.com/google/fully-homomorphic-encryption}{Google FHE}
\href{https://github.com/microsoft/SEAL}{SEAL},
\href{https://github.com/zama-ai/concrete/}{Concrete},
\href{https://github.com/JohnCremona/eclib}{eclib},
\href{https://github.com/homenc/HElib}{HElib}, and
\href{https://gitlab.com/palisade/palisade-release}{PALISADE}.
    \\
    \hline
    
    Trusted execution environments (TEE) 
    &  Hardware and software that provide computation security against the unwarranted retrieval of sensitive information~\cite{OMTP_TEE}. 
    &  \href{https://github.com/Samsung/mTower}{mTower},
\href{https://github.com/openenclave/openenclave}{Open Enclave SDK},
\href{https://docs.nvidia.com/jetson/l4t/index.html#page/Tegra\%20Linux\%20Driver\%20Package\%20Development\%20Guide/trusty.html#}{Trusty},
\href{https://git.trustedfirmware.org/TF-A/trusted-firmware-a.git/}{TrustZone}, 
\href{https://github.com/deislabs/mystikos}{Mystikos},
\href{https://github.com/Open-TEE}{Open-TEE} and
\href{https://software.intel.com/content/www/us/en/develop/articles/intel-trusted-execution-technology-intel-txt-enabling-guide.html#_Toc383534397}{Intel's Trusted Execution Technology}.
    \\
    \hline 
    
    Federated \newline  Learning (FL) 
    & Distributes ML models across data sources for training and averages the weights into one model~\cite{konecny_federated_2015,FL_challenges}.
    & \href{https://fate.fedai.org/}{Fate},
\href{https://github.com/sherpaai/Sherpa.ai-Federated-Learning-Framework}{sherpa.ai},
\href{https://github.com/PaddlePaddle/PaddleFL}{PaddleFL},
 \href{https://github.com/OpenMined/PySyft}{PySft},
\href{https://github.com/xaynetwork/xaynet}{Xaynet},
\href{https://github.com/scaleoutsystems/fedn}{fedn},
\href{https://github.com/FedML-AI}{FedML-AI},
\href{https://github.com/adap/flower}{Flower},
\href{https://github.com/OpenMined/PyVertical}{PyVertical},
\href{https://www.tensorflow.org/federated}{TensorFlow Federated}, and
\href{https://github.com/IBM/federated-learning-lib}{federated-learning-lib}.
    \\
    \hline 

    Blockchain \newline (no PET)
    & Tamper-proof, distributed database, whose state is replicated and stored across P2P network nodes using a consensus algorithm~\cite{butijn2020blockchains}.
    & 
    \href{https://github.com/corda/corda}{Corda}, 
    \href{https://github.com/hyperledger}{Hyperledger}, 
    \href{https://github.com/ethereum/go-ethereum}{Go Ethereum}, 
    \href{https://github.com/bigchaindb}{BigchainDB}, 
    \href{https://github.com/smartcontractkit/chainlink}{Chainlink}, 
    \href{https://github.com/trufflesuite/ganache-ui}{Ganache}, 
    \href{https://github.com/XRPLF/xrpl-dev-portal/}{XRPLF}, 
    \href{https://github.com/cryptonomex}{Graphene}, 
    \href{https://github.com/maticnetwork/}{Polygon}, 
    \href{https://github.com/vechain}{Vechain}, and
    \href{https://gitlab.com/tezos/tezos}{Tezos}.
    \\
    \hline

  \end{tabular}
  
\end{table*}

\endgroup

\begin{table*}[t]
  \centering
  \scriptsize
  \caption{Technology capabilities fulfilling use case characteristics}
  \label{tab:capabilities}
  \renewcommand{\arraystretch}{1.2}
  \begin{tabular}{|p{2.2cm}|p{1.5cm}|p{3.2cm}|p{1.5cm}|p{1.5cm}|p{2.5cm}|p{1.5cm}|}
  \hline 
     \textbf{Technology} 
     & \textbf{Privacy} 
     & \textbf{Function type} 
     & \textbf{Data volume} 
     & \textbf{Data \newline authenticity}
     & \textbf{Query type} 
     & \textbf{Number of \newline interacting parties}
     \\
    \hline
    
    DP
    & Anonymity
    & Noise added to data processing 
    & TB
    & Noisy outputs
    & Known / Unknown
    & One
    \\
    \hline

    K-anonymity
    & Anonymity
    & Dataset  anonymization 
    & GB
    & Generalized
    & Known / Unknown
    & One
    \\
    \hline

    Synthetic data
    & Anonymity
    & Dataset generation 
    & TB
    & Noisy
    & Known / Unknown
    & One
    \\
    \hline
    
    ZKP 
    & Confidentiality
    & Authenticity  proofs
    & MB
    & Yes
    & Known
    & Two
    \\
    \hline
    
    SMC 
    & Confidentiality
    & Arbitrary
    & MB
    & Yes
    & Known
    & Multiple
    \\
    \hline
    
    HE 
    & Confidentiality
    & Arbitrary
    & MB
    & Yes
    & Known
    & Two
    \\
    \hline
   
    TEE 
    & Confidentiality
    & Arbitrary
    & GB
    & Yes
    & Known /  Unknown
    & Multiple
    \\
    \hline   
    
    FL 
    & Confidentiality
    & ML
    & TB
    & Yes
    & Known
    & Multiple
    \\
    \hline
    
    Blockchain  (no PET)
    & Not applicable
    & Arbitrary
    & MB
    & Yes
    & Known
    & Multiple
    \\
    \hline    
    
    \multicolumn{7}{l}{Legend: DP = Differential privacy; ZKP = Zero-knowledge proof; SMC = Secure multiparty computation; HE = Homomorphic encryption;  TEE = Trusted}\\  
    \multicolumn{7}{l}{execution environments; FL = Federated learning} \\   
    
\end{tabular}    

\end{table*}

Table~\ref{tab:use_cases} describes the eigth identified application domains and the use cases in these domain. In this section, we discuss selected application domains in detail, focusing on challenges and opportunities for deploying PETs.  

Recommender systems (\#1 in Table~\ref{tab:use_cases}) can enhance customer experience by suggesting location or automatically activating capabilities, such as the seat heating. However, the data required for such use cases is often highly sensitive. PETs may help reduce the amount of data that needs to be transmitted to centralized clouds while retaining the utility of data-driven recommendations.

Computer vision (\#3 in Table~\ref{tab:use_cases}) utilizes complex machine learning (ML) models to extract information from images and video. However, the unstructured nature of the input data increases the risk of unknowingly capturing sensitive information, e.\,g., people, and drives the need for the usage of PETs, e.\,g., for anonymization data using blurring techniques and synthetic data. The use of federated learning can reduce the need to centralize data and can thus further reduce risks.

% Further, utilizing federated learning, prevent the risks for which privacy regulations were designed, such as data breaches or misappropriation of PII.

Sensitive data management (\#4 in Table~\ref{tab:use_cases}) describes the process of preparing data for secondary purposes. For this purpose, complex and automated data transformation pipelines are required for data anonymization. These pipelines should require only a minimal amount of human intervention. 
% However, companies with manual anonymization processes and data access request pipelines might have human error and delays for analysts to access data. 
PETs, such as k-anonymity and differential privacy, are essential to provide the required privacy guarantees.

% for further processing.a standard that can automatically anonymize datasets for secondary purposes with reduced human intervention.

Data analytics describes the process of using data to support decisions in the business (\#5 in Table~\ref{tab:use_cases}). For this purpose, it is required to aggregate data and connect various data sources. Analytics can be categorized in exploratory, i.\,e., the objective and business question of the analysis is not completely defined yet, and operational analytics, i.\,e., the KPI and business decision is well-specified. As for both types of analytics, it is often unnecessary to expose individual records and all attributes. PETs like k-anonymity and differential privacy can limit the amount of information exposed to analysts. However, there is an important trade-off between utility of the data and privacy to be considered,  in particular, for exploratory analytics.

% objective not well defined) and differenteriate   However, it is not necessary to account for individual items in the data sets. 
% PETs, such as differential privacy
% Furthermore, the processes to request permission for or the anonymization prior to analysis of data containing PII  could be accelerated and enhanced, respectively.
% PETs that allow the execution of intrinsically private functions can enable analysts to perform statistics, e.\,g., with SQL queries, without needing to ``see'' the datasets themselves (differential privacy).

While there is sensitive information that corporations would prefer to maintain private, such as business secrets, performance metrics, or suppliers, cross-organizational data sharing increasingly becomes a necessity to optimize entire value chains and business networks, e.\,g., to support asset search (\#6) and cross-organizational sharing (\#8 in Table~\ref{tab:use_cases}). Asset search addresses the need to locate and track components and products across organizations. Cross-organizational data sharing envisions the sharing of more comprehensive data sets.
PETs can address the need to expose the minimal amount of data and the ability to verify data and results. Emergent platforms, such as GAIA-X~\cite{noauthor_gaia_x_nodate}, heavily rely on PETs to establish secure data exchange mechanisms and controls.

% While anonymization cannot guarantee the exactness, these use cases can benefit from technologies that offer confidentiality.
% For example, using schemata that can detect a component ID within a dataset without revealing these two inputs (secure multiparty computation).

% Furthermore, recent initiatives like GAIA-X~\cite{noauthor_gaia_x_nodate} and the automotive Catena-X encourage sharing data and analytics across enterprises 

% Cross-organizational data sharing (\#~8 in Table~\ref{tab:use_cases}), which would allow companies to access more data to train their algorithms better and potentially extract new insights.
% However, companies may hesitate to share data without architectures that ensure confidentiality.
% Therefore, PETs that protect inputs could unlock collaborations in these initiatives (e.\,g., federated learning or trusted execution environments).

\section{Privacy-Enhancing Technologies: Capabilities and Applications}
\label{sec:pets}

PETs comprise technologies designed to protect the privacy of data owners.
PETs accomplish this by enhancing \emph{anonymity} with technologies such as differential privacy (DP), k-anonymity, or synthetic data, or \emph{confidentiality} with secure and outsourced computation technologies such as zero-knowledge proof (ZKP), secure multi-party computation (SMC), homomorphic encryption (HE), trusted execution environments (TEE), or federated learning (FL).
Furthermore, PETs can provide capabilities for supporting the use cases described in Table~\ref{tab:use_cases}.
Table~\ref{tab:priv_tools} provides an overview of important PETs and the most
relevant open-source tools, which we resulted from our gray literature review. 
% only because it can help to verify claims; without verification, subsequent processing steps may be invalid.
% and is considered a \emph{privacy anti-pattern}~\cite{garrido_revealing_2021}

While blockchain is not strictly a PET, we included it because it is an instrumental building block for establishing trust and support for data verification use cases. Additionally, blockchain can anchor trust of zero-knowledge proof protocols that prove a claim without engaging in sequential messaging~\cite{zkp_anonymous_credentials}.

% \gm{@Andre. Kindly refine why we introduce blockchain.}

\subsection{Characteristics and Capabilities}
\label{subsec:capabilities}

Based on an in-depth analysis of the use cases, we define six important characteristics for selecting PETs and architecting privacy-preserving systems. 

\textbf{Privacy.} This characteristic describes the sensitivity of the data, e.\,g., the need to anonymize personally identifiable information (PII)  and confidential information. Anonymization removes the link between data and individuals. Confidentiality requirements may also exist for non-personal data, e.\,g., due to business reasons.

\textbf{Function type.} Use cases may require the use of analytics queries, ML models, or proofs for the authenticity of data. Depending on the function, PETs, such as basic queries to verify the existence of an asset in a dataset (SMC), aggregation queries (DP), or ML (FL), can be chosen. 

\textbf{Data volume.} Some technologies are more suitable than others, depending on the data volume the use case is processing. 
The noise added by DP is independent of the data volume, while SMC cannot process large data volumes given the encryption and communication overhead. 

\textbf{Data authenticity.} For high-value data, blockchain-based data verification might be necessary to ensure authenticity. Some PETs reduce the authenticity, e.\,g., anonymization perturbs the exact value of data points to disjoint attributes from the users who generated the data. 

% \alnote{how? On the other hand, PETs enhancing confidentiality preserve data authenticity.} 

\textbf{Query type.} Some use cases require exploratory queries (unknown), while others repetitively execute well-defined queries and ML models.
For example, to train an ML model with FL, one must know what the model will predict or classify. A TEE can execute arbitrary user-defined functions (including ML), and transforming a dataset into k-anonymous or synthetic data does not necessarily require knowing in advance the query types. 

\textbf{Number of interacting parties.} Some use cases require data and interactions from more than one entity to interact. For example, FL can train a model distributed across potentially different data owners. SMC jointly computes a function based on the inputs of multiple parties, and DP allows an analyst to query a dataset.

PETs provide different capabilities to address use cases requirements and characteristics. Table~\ref{tab:capabilities} summarizes the capabilities provided by the defined PETs.

% Ultimately, we hope these six characteristics help practitioners map their use cases to concrete technologies. 

\subsection{PETs and automotive use cases}

Understanding use case characteristics and the capabilities of PETs is essential to architect privacy-preserving and practical systems.
Figure~\ref{fig:framework} illustrates our framework for mapping  use cases to  suitable PETs based on the six defining characteristics and capabilities. 
Table~\ref{tab:use_cases_pets} investigates eight automotive use cases and illustrates how a specific PET can address the privacy requirements of each use case. 

The mapping is intended to be illustrative, not complete. It  emphasizes the strengths and weaknesses of the PETs, helping practitioners align PETs and use case requirements. Thus, we have selected reference use cases to highlight the unique benefits of a specific PET. In practice, a combination of PETs is often required to implement a use case end-to-end. We continue with an in-depth discussion of three use cases.

% We investigated (i) the most relevant PETs, and (ii) high-impact use cases derived from real-world industry applications.
% In between the two layers, we investigated a set of technology capabilities amenable to characteristics that practitioners may identify in their use cases.

\begin{figure}
    \centering
    \includegraphics[scale=0.6]{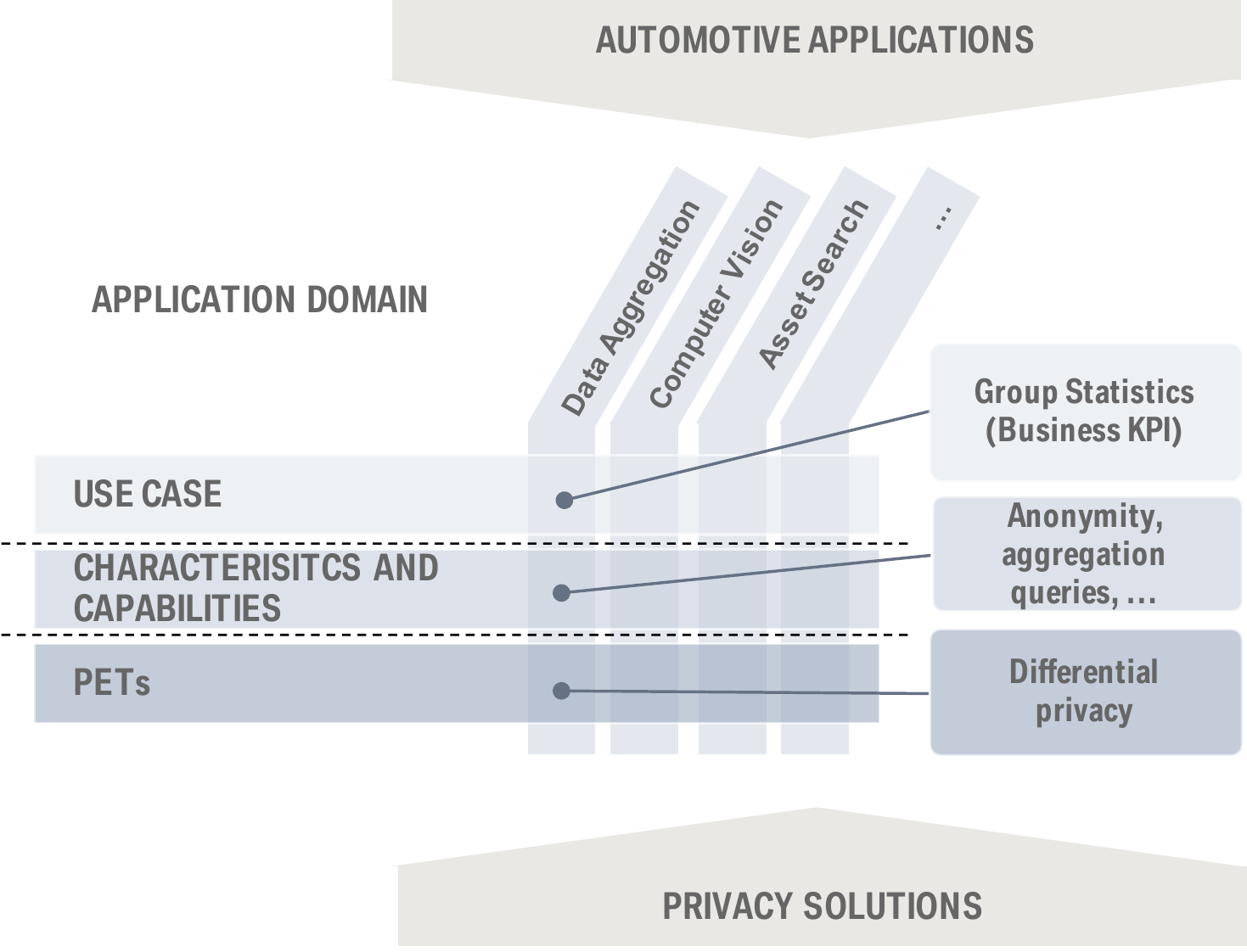}
    \caption{Framework to map use cases and privacy-enhancing technologies.}
    \label{fig:framework}
\end{figure}

\begin{table*}[t]
  \centering
  \scriptsize
  \caption{Reference automotive use cases mapped to privacy-enhancing technologies}
  \label{tab:use_cases_pets}
  \begin{tabular}{|p{0.05cm}|p{3cm}|p{6.5cm}|p{4.5cm}|p{1.5cm}|}
  \hline 
     \textbf{\#} 
     & \textbf{Domain: Use Case} 
     & \textbf{Description} 
     & \textbf{Suitable Capabilities} 
     & \textbf{PETs} \\
    \hline

    1
    & Recommender systems: \newline eco-friendly driving
    & Traing of ML models from complex distributed datasets containing numerous vehicle signals to predict what patterns improve eco-friendly driving.
    & Anonymity, ML over anonymous data, TB of data, noisy data, unknown queries, one party
    & Synthetic \newline data \\
    \hline

    2
    & Geoservices: \newline charging
    & Discovering most frequent locations on an aggregated dataset where electric vehicles have low batteries.
    & Anonymity, aggregation query functions over (anonymous) dataset, GB of data, noisy outputs or generalized data, unknown queries, one party
    & DP,\newline  k-anonymity\\
    \hline

    3
    & Computer vision: \newline attentiveness detection 
    & Training ML models across multiple vehicles and devices.
    & Confidentiality, ML functions, TB of data, authentic data, known queries, multiple parties 
    & FL \\
    \hline
    
    4
    & Sensitive data management: \newline automating anonymization 
    & A practitioner automates the anonymization of ingested customer vehicle data.
    & Anonymity, anonymization, GB of data, generalized data, unknown queries, one party 
    & K-anonymity\\
    \hline
    
    5
    & Data analytics: \newline group statistics 
    & Computing aggregate business KPIs for dashboards by querying various datasets without downloading the underlying data.
    & Anonymity, aggregation query functions, up to TB of data, noisy outputs, unknown queries, one party
    & DP \\
    \hline
    
    6
    & Asset search: \newline tracking components 
    & Tracking components and parts across the value chain to optimize supply chain management (e.\,g., management of stock levels).
    & Confidentiality, arbitrary function, MB of data, authentic data, known query, multiple parties
    & SMC \\
    \hline
    
    7
    &IoT: \newline Connected car
    & Management of vast amounts of sensor data from vehicles and traffic infrastructure across the edge and cloud.
    & Confidentiality, arbitrary functions, MB of data, authentic data, known query, two parties
    & HE \\
    \hline
    
    8
    & Cross-organizational data sharing: 
    Logistics \& supply chain 
    & Track and share data across organizations to optimize business processes, e.\,g., for improved supply chain visibility~\cite{partchain}.
    & Confidentiality, arbitrary functions, GB of data, authentic data, known queries, multiple parties 
    &  TEE, \newline blockchain \newline (anchors trust) \\
    \hline
    
    \multicolumn{5}{l}{Legend: DP = Differential privacy; SMC = Secure multiparty computation; ZKP = Zero-knowledge proof; HE = Homomorphic encryption; TEE = Trusted execution} \\    
    \multicolumn{5}{l}{environments; FL = Federated learning; ML = Machine learning} \\   
    \end{tabular}

\end{table*}

\par{\textbf{Computer vision: attentiveness detection (\#3 in Table~\ref{tab:use_cases_pets}).}}
Alerting drivers of their lack of attention behind the wheel can prevent road accidents and save lives. The training of  ML models typically requires large volumes of potentially sensitive training data. Thus, an important building block is anonymized and synthetic data, particularly for bootstrapping the ML model. However, due to the safety-critical nature, anonymization approaches are not sufficient alone. Federated learning allows the training of models across multiple vehicles without the need of centralizing data, and thus, preserving confidentiality.

% Given the set of capabilities, federated learning is suitable for this use case.

\par{\textbf{Data analytics: group statistics (\#5 in Table~\ref{tab:use_cases_pets})}.} Data warehouses and data lakes are essential enablers for analytics. Data anonymization is an important practice for enabling  secondary data usage. Once datasets are anonymized, an analyst can execute a potentially manifold set of queries, e.\,g., joining and exploring many attributes of vehicles. The use of differential privacy (DP) can prevent the de-identification of data while retaining the utility of the  analysis.  
Using a well-calibrated noise mode a good query accuracy is ensured while preserving each individual's anonymity. DP is also an important enabler for more democratized data access and analytics.
% DP can add noise to preserve anonymity formally, be adapted to aggregate queries such as the count, average, variance, percentiles, among others, and limit their privacy leakage.
Differential privacy can also be applied on the fly, e.\,g., using a DP-aware SQL engine and a privacy budget that controls the number of queries allowed.

% Companies employing SQL to interface with their data warehouses can use DP by deploying a query re-writer that transforms aggregation queries into intrinsically private ones.

\par{\textbf{Asset search: tracking components (\#6 in Table~\ref{tab:use_cases_pets}).}}
The automotive value chain is highly complex, involving many partners in an international network. As a result, supply chains are highly complex. They often lack visibility and trust, in particular concerning tier-n suppliers, i.\,e., suppliers that are not directly in contact with an automotive company. A critical capability is the tracking of components and parts in this cross-organizational network. Blockchains provide a mean to orchestrate a decentral business network~\cite{partchain}. However, additional PETs are essential to facilitate secure data exchange, e.\,g., secure multi-party computation (SMC) enables the secure computation, e.\,g., to reconcile stock levels, avoiding the exposure of confidential business information. However, SMC is only suitable for specific, well-defined use cases, small data volumes, and certain types of computation.

% For example, if data authenticity is not necessary for a study based on sensitive data across many users, a combination of federated learning and differential privacy can bring confidentiality and anonymization together to increase protection. 

An important characteristic of many use cases is data verification and the establishment of trust in distributed and cross-organizational environments. Blockchains and zero-knowledge proofs (ZKP) are an important enablers for these requirements.  They allow the sharing of proofs without revealing the underlying data. For example, individuals can reveal identity-related attributes (e.\,g., the possession of a driver's license~\cite{zkp_driverlicense}) using ZKP.

\section{Related work}
\label{sec:related_work}

Most research focuses on applying or optimizing a single PET to tackle one particular use case, or investigate the use cases that a single PET can address. 
Examples include applying SMC to privacy-preserving deep learning~\cite{bittner2021private}, implementing DP in the context of sensitive health data~\cite{choudhury2020differential}, or identifying applications for which practitioners can employ TEEs~\cite{arfaoui2014}. 
However, these publications do not provide an overview of privacy use cases for different PETs.

Other publications have surveyed how PETs fulfill privacy requirements in general~\cite{heurix2015} or from a particular context such as data exchanges~\cite{pennekamp2019dataflow}. 
Alternatively, publications highlight market opportunities for PETs to solve business problems, e.g., build trust or establish a competitive advantage~\cite{jaatun2012}).
However, mapping PETs with requirements or business opportunities does not provide immediate insights regarding privacy use cases.
Another set of publications proposes industry use cases without explicitly mapping them to a list of PETs.
Examples range from outlining privacy use cases in the supply chain~\cite{gonczol2020}, the role of PETs in predictive maintenance in the automotive industry~\cite{theissler2021}, or the use of PETs in the context of IoT~\cite{data_makets_properties_II} or smart cities~\cite{curzon2019}. 

%\alnote{can we also research some application survey for industrial and maybe other application types?} 
We identified a few publications that survey applications of PETs. There is a repository of implemented PET use cases~\cite{cdei2021} from different sectors (e.\,g., health, transport, finance) and a list of case studies that used PETs to reach their objectives~\cite{ffis2020} in the financial sector. However, these surveys do not focus on production and industry use cases.

While the publications covering the domain of privacy and use cases are varied, to the best of our knowledge, they do not 
%(i) consider pluggable privacy products \alnote{are these products really pluggable} beyond the underlying PETs like those provided by the classified startup vendors,\alnote{I think the first differentiation point does not really apply} 
(i) identify suitable capabilities required by use cases to map them to PETs, (ii) present actionable use cases in the automotive industry, (iii) include a list of reference use case that succinctly demonstrate the value of each PET.

\section{Discussion and Conclusion}
\label{sec:conclusion}

% We encourage practitioners to use our framework in Fig~\ref{fig:framework} to map PETs with use cases based on their characteristics.

% repeat why this paper is important and method
While PETs have matured and are increasingly available, developing privacy-preserving architectures is challenging, requiring an in-depth understanding of PETs and use cases. This paper addresses this challenge and provides guidelines synthesized from  expert interviews and a literature review. 

PETs provide the ability to increase the protection of data while in-use and can be considered complementary to established security practices, e.\,g., security monitoring, data encryption at rest and in-transit, data governance.  There is no ``one-size-fits-all'' privacy-enhancing technology (PET). The selection and deployment of PETs require a careful understanding of use cases characteristics, the capabilities of a PET and its limitation. We demonstrated how use case characteristics can be used to assess the suitability for PETs. While this paper focuses on automotive use cases, the identified characteristics and capabilities generalize well to other application domains. 

The usage of PETs is associated with increased architectural and operational complexity, and performance-related constraints that must be carefully considered when choosing a PET. Further, the limitations of PETs must be carefully considered. For example, homomorphic encryption and secure multi-party computation cannot handle large volumes of data and do not address anonymization requirements, e.\,g., for secondary data processing. K-anonymity does not provide a formal guarantee of privacy like differential privacy. 

%Federated learning is only suitable for ML.

% Furthermore, through the mappings of Tables~\ref{tab:capabilities} and~\ref{tab:use_cases_pets}, we help practitioners understand the landscape of privacy from an industry perspective and swiftly select PETs based on the capabilities amenable with the characteristics of a use case.
% Moreover, we provide a list of key findings beyond the gathered solutions and use cases.
% From these findings we underline the three use case characteristics that practitioners can study to determine the degree of privacy required in a use case, and the importance of engaging in internal ``sensitive data management'' first, otherwise analytics will be hindered and potentially subject to regulation infringements.

% future work
The importance of PETs will increase. In particular, the need to collaborate across organizational boundaries will intensify the need for PETs. In the future, we will refine and extend our classification to other application categories and domains. Further, we implement and experiment with concrete PETs and use cases, e.\,g., differential privacy and secure multi-party computing.

\printbibliography
%\bibliographystyle{unsrt}
%\bibliography{refs}

%  \appendix

\end{document}